\documentclass[aps,prl,reprint,superscriptaddress,floatfix]{revtex4-1}
\usepackage{graphicx}
\usepackage{amsmath,amssymb,bm}
\usepackage{subfigure}
\usepackage[english]{babel}

\newcommand{\li}{\begin{aligned}}
\newcommand{\eli}{\end{aligned}}

\newcommand{\be}{\begin{equation}}
\newcommand{\ee}{\end{equation}}

\begin{document}

\title{Microfluidic Actuation by Einstein--de Haas Spin Torque}

\author{Xin Hu }
\affiliation{%
Kavli Institute for Theoretical Sciences, University of Chinese Academy of Sciences, Beijing, 100190, China.
}%

\author{Mamoru Matsuo }
\email{mamoru@ucas.ac.cn}
\affiliation{%
Kavli Institute for Theoretical Sciences, University of Chinese Academy of Sciences, Beijing, 100190, China.
}%
\affiliation{%
CAS Center for Excellence in Topological Quantum Computation, University of Chinese Academy of Sciences, Beijing 100190, China
}%
\affiliation{%
Advanced Science Research Center, Japan Atomic Energy Agency, Tokai, 319-1195, Japan
}%
\affiliation{%
RIKEN Center for Emergent Matter Science (CEMS), Wako, Saitama 351-0198, Japan
}%

\date{\today}

\begin{abstract}
We propose spin-current microfluidic actuation of a sealed liquid metal.  Spin angular momentum injected from Pt contacts enters the liquid as an Einstein--de Haas torque and is converted through micropolar angular-momentum balance into viscous flow without pressure drive, moving walls, magnetic fields, Lorentz forces, or charge flow through the liquid.  The dc velocity obeys universal spin-diffusion scaling, and the finite-frequency spin-mechanical admittance resolves viscous momentum diffusion, spin transport, microrotation relaxation, and interface transparency of the liquid-metal channel.
\end{abstract}

\maketitle 

\paragraph{Introduction---}
Microfluidics controls fluids in geometries where viscous, surface, and interfacial effects dominate over inertia, and has become a platform for transport, analysis, soft electronics, and thermal management \cite{squires2005microfluidics}.  Conducting liquid metals extend this platform because they remain fluid while retaining metallic electrical and thermal conductivities.  Gallium-based alloys form reconfigurable microfluidic electrodes, sensors, antennas, interconnects, pumps, valves, heaters, and soft electronic components \cite{khoshmanesh2017liquid}.  Their thermal properties make liquid-metal micro/mini-channel heat sinks candidate platforms for high-heat-flux electronics \cite{Chen2025LiquidMetalChipCooling}.  Downsizing liquid-metal fluidics toward integration with micro- and nanoelectromechanical systems makes moving parts, sealing, pressure-driven filling, surface pinning, and interfacial reproducibility increasingly restrictive \cite{LaserSantiago2004Micropumps,khoshmanesh2017liquid}.  A sealed conducting microflow therefore demands a boundary drive that produces a mechanically readable fluid response without pressure-driven operation, moving walls, or charge transport through the liquid.

\begin{figure}[t!]
    \centering
    \includegraphics[scale=0.18]{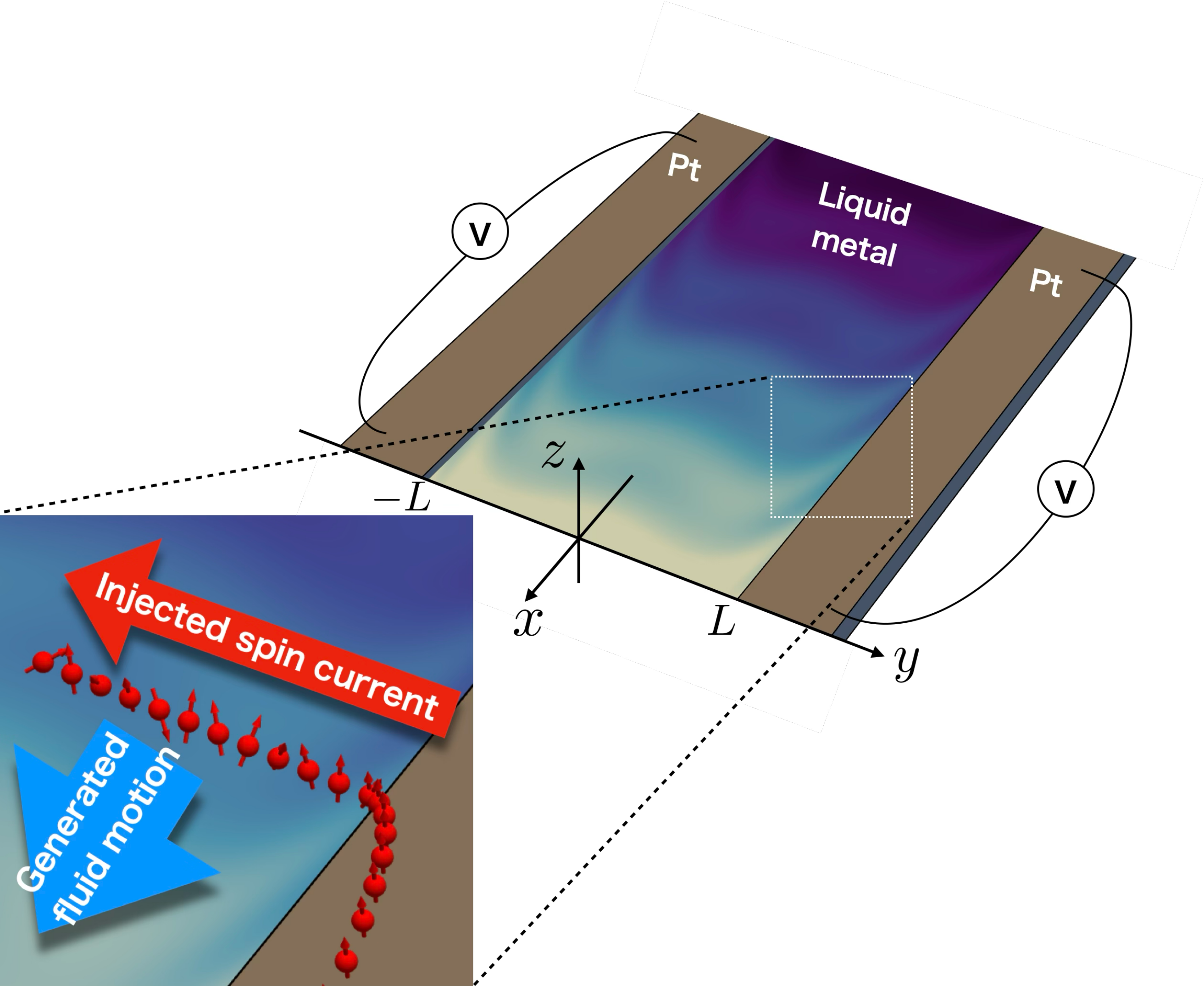}
    \caption{Schematic of the proposed spin-current actuation geometry. A conducting liquid metal is confined in a channel of width $2L$ between two Pt electrodes. A charge current in Pt generates, through the spin Hall effect, a $z$-polarized spin current injected into the liquid metal at $y=\pm L$. The injected spin angular momentum produces an inhomogeneous spin chemical potential and drives liquid-metal motion through the Einstein--de Haas angular-momentum-transfer channel.}
    \label{fig:setup}
\end{figure}

Spin injection supplies such a boundary drive by delivering angular momentum across a solid/liquid-metal interface, where it couples to mechanical rotation rather than to a Lorentz body force.  This contrasts with conventional magnetically driven microfluidics, where electric currents, magnetic fluids, particles, or droplets are manipulated by field-induced forces and torques \cite{Nguyen2012MicroMagnetofluidics}.  The Einstein--de Haas (EdH) effect \cite{einstein1915experimental} and Barnett effect \cite{PhysRev.6.239}, together with gyromagnetic-ratio experiments \cite{RevModPhys.34.102}, establish conversion between spin magnetization and mechanical rotation.  Spin-rotation coupling provides the spintronic form of the same angular-momentum principle \cite{Mashhoon1988,HehlNi1990,PhysRevLett.106.076601,PhysRevB.87.180402}.  Modern gyromagnetic and spin-mechanical experiments now span microcantilevers \cite{wallis2006einstein}, nanomechanical torque detection \cite{Zolfagharkhani2008}, Barnett and gyromagnetic-reversal measurements \cite{Ono2015GyromagneticPRB,Ogata2017GyromagneticAPL,Hirohata2018SpinAccumRotation,Imai2018GyromagneticReversal,Imai2019BarnettFerrimagnet,Chudo2014APEX,Chudo2015JPSJ,Harii2015JJAP,Chudo2021PRB,Chudo2021JPSJ}, nuclear-spin mechanical control \cite{Chudo2025MechanicalMultiplexer}, rotating spin qubits \cite{Wood2017RotatingSpin,Wood2018RotatingSpinQubit,Jin2024RotatingDiamond}, acoustic spin-current generation \cite{Kobayashi2017SAWSpinRotation,Kurimune2020PRB,Kurimune2020PRL,Tateno2020PRB,Tateno2021PRB}, spin-Seebeck mechanical force \cite{Harii2019}, radio-frequency EdH response \cite{Mori2020EdHRadio}, nonreciprocal and gradient-material spin-torque generation \cite{Okano2019Nonreciprocal,Horaguchi2025Gradient}, and ultrafast lattice angular-momentum transfer \cite{Dornes2019}, establishing spin angular momentum as a mechanical control variable across micro- and nanoscale systems.

In liquids, direct spin hydrodynamics has established the reciprocal conversion.  Pressure-driven vorticity generates spin current or spin-dependent voltages in liquid metals \cite{8c8475edb5e04ea7ba7ca651a3a6a117,Takahashi2020,TokoroTakahashi2022}, with theory \cite{PhysRevB.96.020401} and pipe- or duct-flow analyses \cite{PhysRevApplied.14.014002,PhysRevFluids.6.043703}.  Related vorticity-to-spin conversion appears at a different energy scale in relativistic heavy-ion collisions, where global hyperon polarization \cite{osti_1392218,STAR2018GlobalLambda,STAR2021XiOmega}, beam-direction polarization \cite{STAR2019BeamDirection}, and spin-orbital measurements \cite{ALICE2020SpinOrbital} indicate exceptionally large vorticity in the quark--gluon plasma \cite{BecattiniLisa2020}.  Relativistic spin hydrodynamics expresses this physics as angular-momentum exchange between spin density and orbital motion through antisymmetric stress \cite{HattoriHongoHuangMatsuoTaya2019}.  The missing operation is the inverse one, namely whether spin angular momentum injected at a liquid boundary can drive and characterize a sealed conducting liquid without charge flow through the liquid, moving walls, an external magnetic field, or a Lorentz force.

In this Letter, we formulate spin-injection microfluidics for a sealed liquid metal in the Pt/liquid-metal/Pt geometry of Fig.~\ref{fig:setup}.  A charge current in Pt generates a pure spin current through the spin Hall effect and injects it into the liquid metal \cite{Saitoh2006,Takahashi_2008}.  We show how an EdH torque deposits injected electron spin into the rotational sector of a viscous conducting liquid.  Antisymmetric stress then converts that angular momentum into flow, and the spin-accumulation gradient enters the Stokes equation as a hydrodynamic body force.  We derive a spin-mechanical response theory for both dc actuation and ac admittance.  The dc pumping signal obeys a universal geometry-controlled scaling, while the ac admittance provides a width- and frequency-resolved method for extracting viscous, spin-transport, microrotational, and interfacial parameters.  A symmetry-related counterflow channel supplies an angular-momentum-resolved diagnostic.  Together these results establish EdH liquid-metal actuation and admittance spectroscopy as a spin-driven actuator element for spin-current circuitry used in all-spin-logic architectures \cite{BehinAein2010AllSpinLogic,maekawa2017spin}.

\paragraph{Model---}
We formulate the inverse spin-hydrodynamic response within the low-frequency micropolar hydrodynamics of a liquid metal.\footnote{See Supplemental Material for the rotational-sector angular-momentum balance, the low-frequency reduction of the micropolar equations, the transient and finite-frequency response functions, and the pressure-constrained and interface-corrected extensions.}  The variables are the fluid velocity $\mathbf{v}(t,\mathbf{r})$ and the microrotation angular velocity $\boldsymbol{\Omega}(t,\mathbf{r})$, which represents the internal rotational sector that can exchange angular momentum with electron spins.  Global rotational symmetry constrains the angular-momentum density and the exchange between orbital and intrinsic rotation, while rotational viscosity, couple stress, and EdH spin transfer enter as constitutive terms.

For a dissipative incompressible liquid, the linear-momentum and intrinsic angular-momentum balances are
\begin{align}
\rho D_t v_i
&=
-\partial_i p+
\partial_jT_{ij},
\nonumber\\[-1mm]
I D_t\Omega_i
&=
\varepsilon_{ijk}T_{jk}
+
\partial_jC_{ji}
+
\tau_i^{\rm EdH}.
\label{eq:balance-main}
\end{align}
Here $T_{ij}$ is the force stress, $C_{ji}$ is the couple stress, $I$ is the micro-inertia density, and $\tau_i^{\rm EdH}$ is the torque density transferred from the electron spin subsystem to the rotational sector.  For an isotropic liquid, the leading dissipative variable in the rotational sector is the relative rotation $A_i=(\nabla\times\mathbf{v})_i-2\Omega_i$.  In the low-frequency limit, the second line of Eq.~\eqref{eq:balance-main} reduces to $0=2\mu_r[(\nabla\times\mathbf{v})_i-2\Omega_i]+\tau_i^{\rm EdH}$, where $\mu_r$ is the rotational viscosity.  We close the electron-spin sector phenomenologically by $\tau_i^{\rm EdH}=4\alpha\mu_r\mu_i^S$, with $\boldsymbol{\mu}^S$ the spin chemical potential in the liquid metal.  The coefficient $\alpha$ is a low-energy spin-to-rotation angular-momentum-transfer coefficient that encodes band-structure, spin-orbit, and spin-relaxation effects.  The local torque balance then yields $\Omega_i=(\nabla\times\mathbf{v})_i/2+\alpha\mu_i^S$.

For the laminar, low-Reynolds-number response, the convective part of $D_t$ is neglected, while the unsteady term is retained for transient and admittance calculations.  Substitution into the incompressible momentum equation yields the EdH-driven Stokes equation, with $\eta=\mu+\mu_r$,
\begin{align}
\rho\partial_t\mathbf{v}
=
-\nabla p
+
\eta\nabla^2\mathbf{v}
-
2\alpha\mu_r\nabla\times\boldsymbol{\mu}^S.
\label{eq:edh-stokes-main}
\end{align}
Equation~\eqref{eq:edh-stokes-main} is the hydrodynamic inverse channel studied below: a curl of spin accumulation acts as a body force because spin angular momentum is first deposited into the rotational sector and then converted into viscous flow.

\paragraph{Universal spin-diffusion scaling---}
We apply Eq.~\eqref{eq:edh-stokes-main} to the Pt/liquid-metal/Pt channel in Fig.~\ref{fig:setup}.  The liquid metal occupies $-L\leq y\leq L$, the spin accumulation is $\boldsymbol{\mu}^S(y)=\mu_z^S(y)\hat{\mathbf{z}}$, and the generated velocity is $\mathbf{v}(y,t)=v_x(y,t)\hat{\mathbf{x}}$.  The spin sector obeys $d^2\mu_z^S/dy^2=\mu_z^S/\lambda^2$ and $j_y^S=-C_s d\mu_z^S/dy$, where $C_s=\hbar\sigma_0/(4e^2)$.  Let $J_+=j_y^S(L)$ and $J_-=j_y^S(-L)$ be the signed boundary spin currents, and define $J_{\rm o}=(J_- - J_+)/2$ and $J_{\rm e}=(J_+ + J_-)/2$.  The $J_{\rm o}$ channel corresponds to inward spin injection from both interfaces and produces an odd counterflow.  The $J_{\rm e}$ channel corresponds to a through-spin-current configuration and produces an even pumping flow.

With no-slip boundaries $v_x(\pm L,t)=0$, the steady solution of Eq.~\eqref{eq:edh-stokes-main} can be written as
\begin{align}
v_x^{\rm st}(y)
=
V_*
\left[
J_{\rm o}\mathcal{F}_{\rm o}\left(\frac{y}{\lambda};\frac{L}{\lambda}\right)
+
J_{\rm e}\mathcal{F}_{\rm e}\left(\frac{y}{\lambda};\frac{L}{\lambda}\right)
\right].
\label{eq:general-scaling-main}
\end{align}
The material prefactor and the two dimensionless shapes are
\begin{align}
V_*
&=
\frac{2\alpha\mu_r\lambda^2}{\eta C_s}
=
\frac{8\alpha\mu_r e^2\lambda^2}{(\mu+\mu_r)\hbar\sigma_0},
\nonumber\\[-1mm]
\mathcal{F}_{\rm o}(\zeta;\ell)
&=
\frac{\sinh\zeta}{\sinh\ell}
-
\frac{\zeta}{\ell},
\,\,
\mathcal{F}_{\rm e}(\zeta;\ell)
=
1-
\frac{\cosh\zeta}{\cosh\ell},
\label{eq:scaling-definitions-main}
\end{align}
with $\zeta=y/\lambda$ and $\ell=L/\lambda$, where $\lambda$ represents the material-only amplitude.    Equations~\eqref{eq:general-scaling-main} and \eqref{eq:scaling-definitions-main} are the central result: after the velocity is normalized by the spin-injection amplitude and the material prefactor $V_*$, the shape of the EdH-driven flow is controlled only by the geometric ratio $L/\lambda$.

For a reference boundary spin current $J_0^S$, we write $v_0=V_*J_0^S$.  The two pure symmetry responses are $v_{x,{\rm o}}^{\rm st}/v_0=\mathcal{F}_{\rm o}$ and $v_{x,{\rm e}}^{\rm st}/v_0=\mathcal{F}_{\rm e}$.  The even mode has a nonzero cross-section average,
\begin{align}
\frac{\bar{v}_x}{v_0}
=
1-
\frac{\tanh(L/\lambda)}{L/\lambda}.
\label{eq:mean-scaling-main}
\end{align}
The odd mode, in contrast, has zero net mass flux and finite half-channel flow, $Q_{\rm half}/(v_0\lambda)=L/(2\lambda)-\tanh[L/(2\lambda)]$.  It provides a symmetry-resolved control: the net mass flux must vanish, whereas the antisymmetric counterflow response can remain finite.  Equivalently, the first moment $-\rho\int_{-L}^{L}yv_{x,{\rm o}}^{\rm st}(y)dy$ can be finite while the total mass current is zero.\footnote{See Supplemental Material for the first-moment interpretation as an orbital-angular-momentum response and for the odd-channel admittance.}

\begin{figure}[t]
    \centering
    \includegraphics[width=0.95\columnwidth]{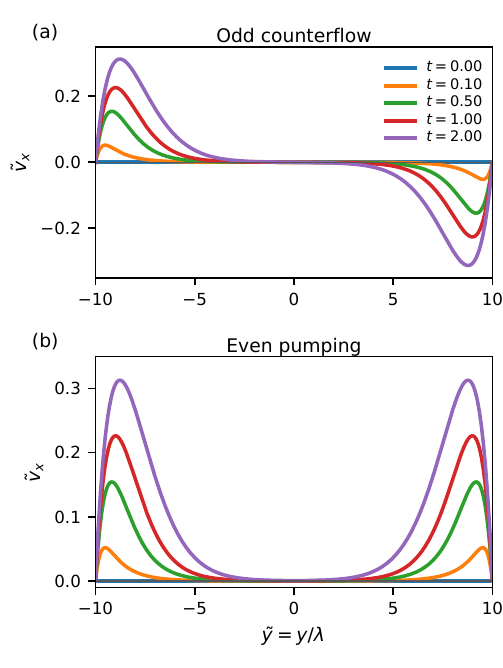}
    \caption{Time evolution of the dimensionless EdH-driven velocity profile $\tilde v_x=v_x/v_0$ at $L/\lambda=10$ for (a) the odd counterflow and (b) the even pumping channel.  The coordinate is $\tilde y=y/\lambda$, and time is measured in units of $\lambda^2/\nu$ with $\nu=(\mu+\mu_r)/\rho$.}
    \label{fig:velocity}
\end{figure}

The transient part is controlled by viscous momentum diffusion.  For the ideal no-slip channel, the slowest relaxation times are $\tau_{\nu}^{\rm o}=L^2/(\pi^2\nu)$ and $\tau_{\nu}^{\rm e}=4L^2/(\pi^2\nu)$, with $\nu=(\mu+\mu_r)/\rho$.  For $L/\lambda=10$, these become $\tau_{\nu}^{\rm o}\simeq10.1\lambda^2/\nu$ and $\tau_{\nu}^{\rm e}\simeq40.5\lambda^2/\nu$.  The time window in Fig.~\ref{fig:velocity} is therefore a build-up interval rather than the final steady saturation.

\begin{figure*}[t]
    \centering
    \includegraphics[width=0.92\textwidth]{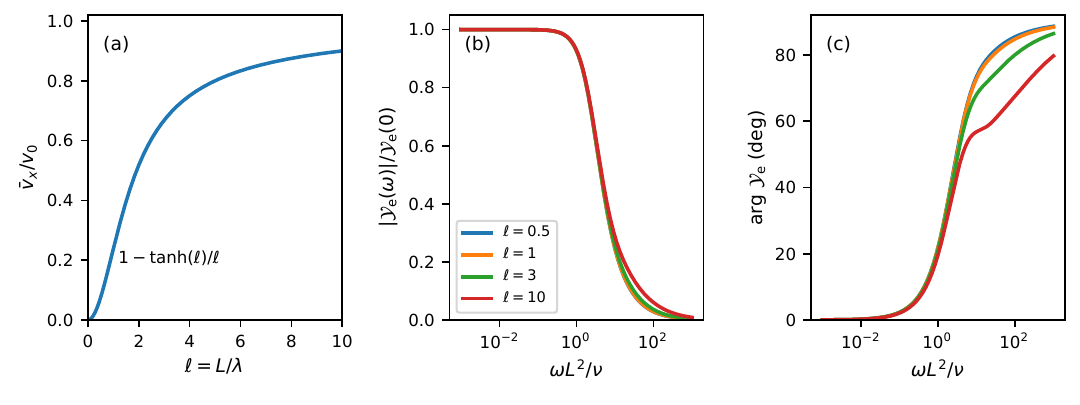}
    \caption{Spin-mechanical response functions for actuation and parameter extraction.  (a) Universal dc pumping response $\bar v_x/v_0=1-\tanh(\ell)/\ell$ with $\ell=L/\lambda$.  (b) Magnitude and (c) phase of the normalized even-channel admittance $\mathcal{Y}_{\rm e}(\omega)/\mathcal{Y}_{\rm e}(0)$ for $L/\lambda=0.5,1,3,10$.  The even channel behaves as a viscous low-pass filter: the liquid follows the spin torque quasistatically for $\omega L^2/\nu\ll1$, whereas the mean-flow amplitude rolls off and the phase approaches quadrature when viscous momentum diffusion cannot span the channel within one period.  Changing the channel width separates the cutoff scale $\nu/L^2$ from the spin-to-flow amplitude.}
    \label{fig:spin-mechanical-response}
\end{figure*}

\paragraph{Spin-mechanical admittance---}
A time-modulated spin current converts the even pumping channel into a spin-mechanical admittance.  Let the injected boundary spin current be modulated as $J_{\rm o,e}(t)=J_{\rm o,e}(\omega)e^{-i\omega t}$ and define $\mathcal{Y}_{\rm e}(\omega,L)=\bar v_{x,\rm e}(\omega)/J_{\rm e}(\omega)$ for the even pumping mode.  When the spin accumulation is quasistatic, the spatial spin-diffusion profile is fixed and the mechanical part of the response is a weighted sum of viscous low-pass filters,
\begin{align}
\mathcal{Y}_{\rm e}(\omega,L)
=
V_*
\sum_{n=0}^{\infty}
\frac{B_n^{\rm e}(L/\lambda)}{1-i\omega\tau_n^{\rm e}},
\,\,
\tau_n^{\rm e}
=
\frac{4L^2}{\nu(2n+1)^2\pi^2}.
\label{eq:ac-admittance-main}
\end{align}
The relaxation times $\tau_n^{\rm e}$ are the viscous momentum-diffusion times of the channel modes; each term therefore has a cutoff set by $\tau_n^{\rm e}$.  The dimensionless weights $B_n^{\rm e}$ are fixed by the spin-diffusion profile.\footnote{See Supplemental Material for the mode expansion and the explicit weights $B_n^{\rm e}(L/\lambda)$.}  At $\omega=0$, Eq.~\eqref{eq:ac-admittance-main} reduces to the dc mean-flow response in Eq.~\eqref{eq:mean-scaling-main}, which is plotted in Fig.~\ref{fig:spin-mechanical-response}(a).  At finite frequency, the admittance becomes a low-pass response.  Its magnitude and phase, shown in Figs.~\ref{fig:spin-mechanical-response}(b) and \ref{fig:spin-mechanical-response}(c), roll off when $\omega L^2/\nu$ is of order unity.  Because the pole positions scale as $L^{-2}$, a width-dependent frequency sweep separates the viscous kinematic coefficient $\nu$ from the overall spin-to-flow amplitude.  For Ga-based liquid metals, representative values of $\nu$ are of order $10^{-7}\,{\rm m^2 s^{-1}}$ \cite{khoshmanesh2017liquid}.  The first even-mode roll-off $f_{\nu}^{\rm e}=1/(2\pi\tau_{\nu}^{\rm e})=\pi\nu/(8L^2)$ then lies in the Hz--kHz range for $L=10$--$100\,\mu{\rm m}$.  Such dimensions are compatible with a pre-confined Pt/liquid-metal/Pt channel or sealed thin slab, where submicron pressure-driven liquid-metal flow is unnecessary.

At higher frequencies the spin and rotational sectors add distinct filters.  Dynamic spin diffusion replaces the static length by
\begin{align}
\lambda_\omega
=
\frac{\lambda}{\sqrt{1-i\omega\tau_s}},
\,\,
\lambda^2=D_s\tau_s,
\label{eq:lambda-omega-main}
\end{align}
and the force profile itself changes when $\omega\tau_s$ becomes appreciable.  Light liquid metals such as Ga are favorable candidates because their weaker spin-orbit coupling can enlarge the spin-diffusion length, although $\lambda$ and $\tau_s$ should be treated as experimentally extracted parameters.  Keeping the finite response of the microrotation sector introduces another dispersive channel,
\begin{align}
\Omega_i(\mathbf{q},\omega)
=
\frac{\frac{1}{2}i(\mathbf{q}\times\mathbf{v})_i+\alpha\mu_i^S(\mathbf{q},\omega)}{1-i\omega\tau_R+\xi^2q^2},
\label{eq:rotational-filter-main}
\end{align}
where $\tau_R=I/(4\mu_r)$ and $\xi$ is the rotational diffusion length associated with couple stress.  Equations~\eqref{eq:ac-admittance-main}--\eqref{eq:rotational-filter-main} show how finite-frequency spin injection turns the channel into a spin-mechanical admittance spectrometer: varying $\omega$ and $L$ can distinguish viscous momentum diffusion, spin diffusion, microrotation relaxation, and interface transparency.

\pagebreak[4]
Figure~\ref{fig:spin-mechanical-response} displays this separation between dc gain and viscous filtering.  Panel (a) shows that the dc mean flow is suppressed as $\bar v_x/v_0\simeq (L/\lambda)^2/3$ when the channel is narrower than the spin-diffusion length, and approaches $\bar v_x/v_0\simeq1$ when $L\gg\lambda$.  Panels (b) and (c) are the frequency response of the same actuator.  For $\omega L^2/\nu\ll1$, viscous momentum diffusion equilibrates the whole channel within each period; the mean velocity follows the injected spin current with constant magnitude and nearly zero phase shift.  For $\omega L^2/\nu\gtrsim1$, momentum cannot diffuse across the channel during one cycle; the liquid therefore acts as a mechanical low-pass filter, suppressing the mean flow and driving the phase toward quadrature.  The ratio $L/\lambda$ changes the modal weights set by the spin-diffusion profile, while the roll-off variable $\omega L^2/\nu$ identifies the viscous time scale.

\paragraph{Experimental signatures and operating window---}
Sign, parity, and harmonic content isolate the EdH response experimentally.  Reversing the injector spin polarization reverses the EdH-driven velocity; in the Pt implementation, reversing the Pt current produces this sign change, and an injector with the opposite spin Hall angle reverses it at fixed charge-current direction.  Switching from the $J_{\rm e}$ channel to the $J_{\rm o}$ channel changes net pumping into odd counterflow: the odd channel has zero net mass flux, finite opposite half-channel flows, and an origin-independent first mass-current moment.  The EdH torque is proportional to the injected spin current, changes sign under injector-current reversal, and appears in the first-harmonic velocity signal.  Joule heating scales as the square of the injector current, remains even under current reversal, and contributes mainly to dc thermal drift or a $2\omega$ response.  A magnetohydrodynamic Lorentz force would require a charge current inside the liquid and an applied magnetic field; both are absent in the ideal spin-current boundary-drive geometry.

The operating window is a low-Reynolds-number confined-response experiment, not pressure-driven injection of liquid metal through a submicron pipe.  The length entering Eqs.~\eqref{eq:mean-scaling-main} and \eqref{eq:ac-admittance-main} is the half-width $L$ of a pre-confined metallic slab between spin-injecting walls, as in Fig.~\ref{fig:setup}.  Capillarity, wetting, and oxide-skin rheology control fabrication and interface reproducibility in gallium-based microfluidics \cite{khoshmanesh2017liquid}; after confinement, the spin-driven measurement reads out the small mechanical response of the sealed conductor.  For $L=10$--$100\,\mu{\rm m}$ and Ga-based values of $\nu$, Eq.~\eqref{eq:ac-admittance-main} places the first even-mode roll-off from the kHz range down to the Hz range, accessible to lock-in detection.  The dc width dependence extracts the effective spin-diffusion length from the universal $L/\lambda$ curve without an absolute spin-to-flow calibration.  The finite-frequency roll-off extracts $\nu$ from the $L^{-2}$ scaling of the admittance poles.  After these two scalings are fixed, the remaining amplitude and systematic deviations diagnose interface transparency, the EdH transfer coefficient, boundary slip, spin-memory loss, oxide-skin pinning, or rotational anchoring.

\paragraph{Conclusion---}
We have formulated EdH spin-torque actuation of a sealed liquid-metal microflow.  In a micropolar liquid, injected spin accumulation transfers angular momentum to the rotational sector, and antisymmetric stress converts that torque into a Stokes body force.  The drive therefore differs from Lorentz-force, magnetic-fluid, pressure, and moving-wall mechanisms because it requires only spin-current injection at the boundary.  For a Pt/liquid-metal/Pt channel, spin diffusion fixes the force profile.  The even boundary condition produces net pumping with universal scaling in channel width relative to the spin-diffusion length, whereas the odd boundary condition yields counterflow with zero net mass flux.  The finite-frequency spin-mechanical admittance supplies the readout.  Its width-dependent roll-off measures viscous momentum diffusion, while additional filters encode spin transport, microrotation relaxation, and interfacial transparency.  Sign reversal, boundary-symmetry selection, and harmonic content separate the EdH response from heating and electromagnetic artifacts.  The boundary spin current is therefore a mechanical control input for sealed liquid-metal microfluidics.  These results establish a spin-driven liquid-metal actuator element for spin-current circuitry used in all-spin-logic architectures \cite{BehinAein2010AllSpinLogic,maekawa2017spin}, without moving walls, external magnetic fields, or charge flow through the liquid.

\paragraph{Acknowledgments---}
M. M. is deeply grateful to J. Kishine for drawing our attention to the close relationship between our spin-mechatronics project and Eringen's micropolar theory. The authors thank H. Chudo, H. Funaki, and T. Funato for valuable discussions.
This work was supported by the National Natural Science Foundation of China (NSFC) under Grant No. 12374126, 
by the Priority Program of Chinese Academy of Sciences under Grant No. XDB28000000, 
and by JSPS KAKENHI for Grants (Nos. 23H01839 and 24H00322) from MEXT, Japan. 

\bibliography{ref}

\clearpage
\onecolumngrid
\setcounter{equation}{0}
\setcounter{figure}{0}
\setcounter{table}{0}
\setcounter{section}{0}
\setcounter{subsection}{0}
\renewcommand{\theequation}{S\arabic{equation}}
\renewcommand{\thefigure}{S\arabic{figure}}
\renewcommand{\thetable}{S\arabic{table}}
\renewcommand{\thesection}{S\Roman{section}}
\renewcommand{\thesubsection}{S\Roman{section}.\Alph{subsection}}

\begin{center}
{\large \textbf{Supplemental Material for\\
Microfluidic Actuation by Einstein--de Haas Spin Torque}}
\end{center}
\section{Micropolar balance}
\label{sec:micropolar-angular-balance}

\subsection{Micropolar variables and angular-momentum density}
\label{subsec:micropolar-variables}

A micropolar liquid assigns to each fluid element both a translational velocity and an intrinsic angular velocity.  We write these fields as
\begin{align}
\mathbf{v}(t,\mathbf{r}),
\,\,
\boldsymbol{\Omega}(t,\mathbf{r}),
\label{eq:micropolar-fields}
\end{align}
with material derivative
\begin{align}
D_t
=
\partial_t+\mathbf{v}\cdot\nabla.
\label{eq:material-derivative}
\end{align}
The intrinsic angular-momentum density of the rotational sector is $I\Omega_i$, where $I$ is the micro-inertia density.  The liquid angular-momentum density is therefore the sum of the orbital angular momentum of the mass current and the intrinsic microrotation contribution,
\begin{align}
\mathcal{J}_i^{\mathrm{liq}}
=
\epsilon_{ijk}r_j\rho v_k+I\Omega_i .
\label{eq:liquid-angular-momentum-density}
\end{align}
If the electron spin density $s_i$ is included, the total angular-momentum density of the liquid plus electron-spin subsystem is
\begin{align}
\mathcal{J}_i^{\mathrm{tot}}
=
\epsilon_{ijk}r_j\rho v_k+I\Omega_i+s_i .
\label{eq:total-angular-momentum-density}
\end{align}
The EdH torque is an internal exchange term between $s_i$ and the liquid rotational sector.  It enters the liquid angular-momentum balance with the opposite sign to its appearance in the electron-spin balance.  Schematically,
\begin{align}
\partial_t\left(\epsilon_{ijk}r_j\rho v_k+I\Omega_i\right)
+\partial_l\mathcal{M}^{\mathrm{liq}}_{li}
&=
\tau_i^{\mathrm{EdH}}+\tau_i^{\mathrm{wall}},
\nonumber\\[-1mm]
\partial_t s_i+
\partial_lJ^s_{li}
&=
-\tau_i^{\mathrm{EdH}}+
\Gamma_i^{\mathrm{spin}}.
\label{eq:liquid-spin-angular-momentum-balance}
\end{align}
Here $\mathcal{M}^{\mathrm{liq}}_{li}$ denotes the liquid angular-momentum flux, including the orbital flux and the couple-stress flux; $J^s_{li}$ is the spin-current tensor; $\tau_i^{\mathrm{wall}}$ is a wall torque; and $\Gamma_i^{\mathrm{spin}}$ denotes spin relaxation or boundary loss.  Thus the EdH torque redistributes angular momentum between the electron-spin and liquid rotational sectors, while wall torques and spin loss account for angular momentum exchanged with the environment.

\subsection{Balance equations and constitutive terms}
\label{subsec:micropolar-balance-edh-limit}

The standard micropolar balance laws used in the main text are
\begin{align}
\rho D_t v_i
&=
-\partial_i p+
\partial_jT_{ij},
\label{eq:linear-momentum-balance}\\[-1mm]
I D_t\Omega_i
&=
\epsilon_{ijk}T_{jk}+
\partial_jC_{ji}+\tau_i^{\mathrm{EdH}}.
\label{eq:intrinsic-angular-momentum-balance}
\end{align}
Here $T_{ij}$ is the force stress and $C_{ji}$ is the couple stress.  The antisymmetric part of $T_{ij}$ transfers angular momentum between orbital motion and intrinsic microrotation; $C_{ji}$ transports intrinsic angular momentum; and $\tau_i^{\mathrm{EdH}}$ is the body couple transferred from electron spin.  With the stress-index convention of Eqs.~\ref{eq:linear-momentum-balance} and \ref{eq:intrinsic-angular-momentum-balance}, the force-stress torque term in the second equation is the counterpart of the torque appearing in the orbital part of Eq.~\ref{eq:liquid-angular-momentum-density}.  Hence, in the absence of wall torques, spin loss, and external body couples, the orbital plus intrinsic angular momentum is conserved.

For an isotropic liquid, the leading dissipative variable in the rotational sector is the relative rotation
\begin{align}
A_i
=
(\nabla\times\mathbf{v})_i-2\Omega_i .
\label{eq:relative-rotation-variable}
\end{align}
The corresponding rotational-viscous torque is proportional to $A_i$.  In the low-frequency limit used for the dc scaling theory, we neglect $ID_t\Omega_i$ and the couple-stress divergence in Eq.~\ref{eq:intrinsic-angular-momentum-balance}.  The local torque balance then becomes
\begin{align}
0
=
2\mu_r\left[(\nabla\times\mathbf{v})_i-2\Omega_i\right]
+
\tau_i^{\mathrm{EdH}},
\label{eq:local-torque-balance-before-closure}
\end{align}
where $\mu_r$ is the rotational viscosity.

The low-energy EdH spin-transfer closure used in the main text is
\begin{align}
\tau_i^{\mathrm{EdH}}
=
4\alpha\mu_r\mu_i^S.
\label{eq:edh-torque-closure}
\end{align}
The spin chemical potential in the liquid metal is $\boldsymbol{\mu}^{S}$, and $\alpha$ is a phenomenological spin-to-rotation angular-momentum-transfer coefficient.  Equations~\ref{eq:local-torque-balance-before-closure} and \ref{eq:edh-torque-closure} yield
\begin{align}
\Omega_i
=
\frac{1}{2}(\nabla\times\mathbf{v})_i+
\alpha\mu_i^S.
\label{eq:microrotation-constitutive-low-frequency}
\end{align}
Substitution into the incompressible momentum equation yields the EdH-driven Stokes equation
\begin{align}
\rho\frac{\partial\mathbf{v}}{\partial t}
=
-\nabla p+
\eta\nabla^2\mathbf{v}
-2\alpha\mu_r\nabla\times\boldsymbol{\mu}^{S},
\,\,
\eta=\mu+\mu_r .
\label{eq:edh-driven-stokes-equation}
\end{align}
The effective kinematic viscosity is
\begin{align}
\nu
=
\frac{\eta}{\rho}
=
\frac{\mu+\mu_r}{\rho}.
\label{eq:effective-kinematic-viscosity}
\end{align}
Equation~\ref{eq:edh-driven-stokes-equation} is the only hydrodynamic equation needed for the ideal dc scaling laws.  Finite microrotation inertia and couple stress are restored in Sec.~\ref{sec:finite-frequency-admittance} when discussing high-frequency filters and rotational boundary conditions.
All channel solutions below use the laminar, low-Reynolds-number limit of this equation: the convective term is absent, while the unsteady term is retained for the transient and finite-frequency responses.

\section{One-dimensional channel geometry}
\label{sec:one-dimensional-channel-geometry}

The channel occupies
\begin{align}
-L\leq y\leq L.
\label{eq:channel-domain}
\end{align}
The injected spin current is polarized along $z$, and the generated velocity is along $x$. We therefore write
\begin{align}
\boldsymbol{\mu}^{S}(y)
=
\mu_z^S(y)\hat{\mathbf{z}},
\,\,
\mathbf{v}(y,t)
=
v_x(y,t)\hat{\mathbf{x}}.
\label{eq:one-dimensional-fields}
\end{align}
Then
\begin{align}
\nabla\times\boldsymbol{\mu}^{S}
=
\frac{d\mu_z^S}{dy}\hat{\mathbf{x}},
\label{eq:one-dimensional-curl}
\end{align}
and Eq.~\ref{eq:edh-driven-stokes-equation} becomes
\begin{align}
\rho\frac{\partial v_x}{\partial t}
=
\eta\frac{\partial^2v_x}{\partial y^2}
-2\alpha\mu_r\frac{d\mu_z^S}{dy}.
\label{eq:one-dimensional-stokes-equation}
\end{align}
We impose the no-slip condition
\begin{align}
v_x(-L,t)=0,
\,\,
v_x(L,t)=0.
\label{eq:no-slip-condition}
\end{align}
The dimensionless variables are
\begin{align}
\zeta
=
\frac{y}{\lambda},
\,\,
\ell
=
\frac{L}{\lambda},
\,\,
s
=
\frac{\nu t}{\lambda^2}.
\label{eq:dimensionless-variables}
\end{align}

\section{Spin diffusion and boundary-current decomposition}
\label{sec:spin-diffusion-boundary-decomposition}

The spin chemical potential obeys
\begin{align}
\frac{d^2\mu_z^S}{dy^2}
=
\frac{1}{\lambda^2}\mu_z^S.
\label{eq:spin-diffusion-equation}
\end{align}
The $z$-polarized spin-current density along $y$ is
\begin{align}
j_y^S(y)
=
-C_s\frac{d\mu_z^S}{dy},
\,\,
C_s
=
\frac{\hbar\sigma_0}{4e^2},
\label{eq:spin-current-definition}
\end{align}
where $\sigma_0$ is the electrical conductivity of the liquid metal. We denote the signed boundary spin currents by
\begin{align}
J_+
=
j_y^S(L),
\,\,
J_-
=
j_y^S(-L).
\label{eq:boundary-current-definition}
\end{align}
The two symmetry channels are
\begin{align}
J_{\mathrm{o}}
=
\frac{J_- - J_+}{2},
\,\,
J_{\mathrm{e}}
=
\frac{J_+ + J_-}{2}.
\label{eq:odd-even-boundary-currents}
\end{align}
The component $J_{\mathrm{o}}$ generates an odd velocity profile and a counterflow. The component $J_{\mathrm{e}}$ generates an even velocity profile and a net pumping response.

Solving Eq.~\ref{eq:spin-diffusion-equation} with Eq.~\ref{eq:boundary-current-definition}, we obtain
\begin{align}
\mu_z^S(y)
=
\frac{\lambda}{C_s}
\left[
J_{\mathrm{o}}\frac{\cosh\zeta}{\sinh\ell}
-
J_{\mathrm{e}}\frac{\sinh\zeta}{\cosh\ell}
\right].
\label{eq:general-spin-potential}
\end{align}
The derivative entering Eq.~\ref{eq:one-dimensional-stokes-equation} is
\begin{align}
\frac{d\mu_z^S}{dy}
=
\frac{1}{C_s}
\left[
J_{\mathrm{o}}\frac{\sinh\zeta}{\sinh\ell}
-
J_{\mathrm{e}}\frac{\cosh\zeta}{\cosh\ell}
\right].
\label{eq:spin-potential-gradient}
\end{align}
The inward-injection configuration corresponds to
\begin{align}
J_+=-J_0^S,
\,\,
J_-=J_0^S,
\,\,
J_{\mathrm{o}}=J_0^S,
\,\,
J_{\mathrm{e}}=0.
\label{eq:inward-injection-current}
\end{align}
The through-spin-current configuration corresponds to
\begin{align}
J_+=J_0^S,
\,\,
J_-=J_0^S,
\,\,
J_{\mathrm{o}}=0,
\,\,
J_{\mathrm{e}}=J_0^S.
\label{eq:through-current-boundary-condition}
\end{align}

\section{Universal steady velocity profiles}
\label{sec:universal-steady-velocity-profiles}

The steady equation is
\begin{align}
0
=
\eta\frac{d^2v_x^{\mathrm{st}}}{dy^2}
-2\alpha\mu_r\frac{d\mu_z^S}{dy}.
\label{eq:steady-stokes-equation}
\end{align}
Define the velocity-response coefficient
\begin{align}
V_*
=
\frac{2\alpha\mu_r\lambda^2}{\eta C_s}
=
\frac{8\alpha\mu_r e^2\lambda^2}{(\mu+\mu_r)\hbar\sigma_0}.
\label{eq:velocity-response-coefficient}
\end{align}
The no-slip solution is
\begin{align}
v_x^{\mathrm{st}}(y)
=
V_*
\left[
J_{\mathrm{o}}\mathcal{F}_{\mathrm{o}}(\zeta;\ell)
+
J_{\mathrm{e}}\mathcal{F}_{\mathrm{e}}(\zeta;\ell)
\right],
\label{eq:general-steady-velocity}
\end{align}
where
\begin{align}
\mathcal{F}_{\mathrm{o}}(\zeta;\ell)
=
\frac{\sinh\zeta}{\sinh\ell}
-
\frac{\zeta}{\ell},
\,\,
\mathcal{F}_{\mathrm{e}}(\zeta;\ell)
=
1-
\frac{\cosh\zeta}{\cosh\ell}.
\label{eq:universal-shape-functions}
\end{align}
The parity and boundary conditions are
\begin{align}
\mathcal{F}_{\mathrm{o}}(-\zeta;\ell)
=
-\mathcal{F}_{\mathrm{o}}(\zeta;\ell),
\,\,
\mathcal{F}_{\mathrm{e}}(-\zeta;\ell)
=
\mathcal{F}_{\mathrm{e}}(\zeta;\ell),
\label{eq:shape-function-parity}
\end{align}
\begin{align}
\mathcal{F}_{\mathrm{o}}(\pm\ell;\ell)
=
0,
\,\,
\mathcal{F}_{\mathrm{e}}(\pm\ell;\ell)
=
0.
\label{eq:shape-function-boundary-values}
\end{align}
Equations \ref{eq:general-steady-velocity} and \ref{eq:universal-shape-functions} are the one-parameter geometric scaling law. All material parameters and spin-injection amplitudes enter through $V_*J_{\mathrm{o}}$ and $V_*J_{\mathrm{e}}$. The normalized shape depends only on $\ell=L/\lambda$.

For a reference spin-current amplitude $J_0^S$, the velocity scale is
\begin{align}
v_0
=
V_*J_0^S
=
\frac{8\alpha\mu_r e^2J_0^S\lambda^2}{(\mu+\mu_r)\hbar\sigma_0}.
\label{eq:velocity-scale-v0}
\end{align}
The two pure symmetry responses are therefore
\begin{align}
\frac{v_{x,\mathrm{o}}^{\mathrm{st}}(y)}{v_0}
=
\mathcal{F}_{\mathrm{o}}\left(\frac{y}{\lambda};\frac{L}{\lambda}\right),
\,\,
\frac{v_{x,\mathrm{e}}^{\mathrm{st}}(y)}{v_0}
=
\mathcal{F}_{\mathrm{e}}\left(\frac{y}{\lambda};\frac{L}{\lambda}\right).
\label{eq:normalized-steady-profiles}
\end{align}

\section{Integral observables}
\label{sec:integral-observables}

The even mode produces a nonzero cross-section average,
\begin{align}
\bar{v}_x
=
\frac{1}{2L}\int_{-L}^{L}v_x^{\mathrm{st}}(y)dy
=
V_*J_{\mathrm{e}}\mathcal{A}_{\mathrm{e}}(\ell),
\label{eq:mean-velocity-definition}
\end{align}
with
\begin{align}
\mathcal{A}_{\mathrm{e}}(\ell)
=
1-
\frac{\tanh\ell}{\ell}.
\label{eq:even-mean-velocity-function}
\end{align}
The volume flow rate per unit length in the $z$ direction is
\begin{align}
Q_x
=
\int_{-L}^{L}v_x^{\mathrm{st}}(y)dy
=
2LV_*J_{\mathrm{e}}
\left(1-\frac{\tanh\ell}{\ell}\right).
\label{eq:even-volume-flow-rate}
\end{align}

The odd mode has zero total flow rate,
\begin{align}
\int_{-L}^{L}v_{x,\mathrm{o}}^{\mathrm{st}}(y)dy
=
0,
\label{eq:odd-zero-total-flow}
\end{align}
whereas its half-channel flow rate is finite. For $J_{\mathrm{o}}>0$,
\begin{align}
Q_{\mathrm{half}}
=
\int_{-L}^{0}v_x^{\mathrm{st}}(y)dy
=
V_*J_{\mathrm{o}}\lambda\mathcal{A}_{\mathrm{o}}(\ell),
\label{eq:half-channel-flow-rate}
\end{align}
where
\begin{align}
\mathcal{A}_{\mathrm{o}}(\ell)
=
\frac{\ell}{2}
-
\tanh\frac{\ell}{2}.
\label{eq:odd-half-flow-function}
\end{align}
The opposite half of the channel carries the same magnitude of flow in the opposite direction.

The odd counterflow has a nonzero orbital angular momentum about the channel center. Per unit length in the $z$ direction, define
\begin{align}
\mathcal{J}_{z}^{\mathrm{orb}}
=
-\rho\int_{-L}^{L}y v_{x,\mathrm{o}}^{\mathrm{st}}(y)dy.
\label{eq:orbital-angular-momentum-definition}
\end{align}
Then
\begin{align}
\frac{\mathcal{J}_{z}^{\mathrm{orb}}}{2\rho V_*J_{\mathrm{o}}\lambda^2}
=
\frac{\ell^2}{3}+1-\ell\coth\ell.
\label{eq:orbital-angular-momentum-scaling}
\end{align}
This observable is independent of the choice of transverse origin within the odd sector. If the origin is shifted by $y_0$, the first moment becomes
\begin{align}
\mathcal{J}_{z}^{\mathrm{orb}}(y_0)
=
-\rho\int_{-L}^{L}(y-y_0)v_{x,\mathrm{o}}^{\mathrm{st}}(y)dy
=
\mathcal{J}_{z}^{\mathrm{orb}}
+
y_0\rho\int_{-L}^{L}v_{x,\mathrm{o}}^{\mathrm{st}}(y)dy.
\label{eq:origin-shift-orbital-angular-momentum}
\end{align}
Using Eq.~\ref{eq:odd-zero-total-flow}, the second term vanishes. Thus the two boundary-current symmetry channels select two distinct hydrodynamic responses: an even mass-pumping mode and an odd origin-independent orbital-angular-momentum mode.

The maximum velocity of the even mode occurs at $y=0$ and is
\begin{align}
\frac{v_{\mathrm{max},\mathrm{e}}}{V_*J_{\mathrm{e}}}
=
1-\operatorname{sech}\ell.
\label{eq:maximum-even-velocity}
\end{align}
For the odd mode, the extremum position is determined by
\begin{align}
\cosh\zeta_*
=
\frac{\sinh\ell}{\ell},
\,\,
\zeta_*
=
\operatorname{arcosh}\left(\frac{\sinh\ell}{\ell}\right).
\label{eq:odd-extremum-position}
\end{align}
The maximum absolute velocity is
\begin{align}
\frac{v_{\mathrm{max},\mathrm{o}}}{V_*J_{\mathrm{o}}}
=
\frac{1}{\ell}
\left[
\operatorname{arcosh}\left(\frac{\sinh\ell}{\ell}\right)
-
\sqrt{1-\left(\frac{\ell}{\sinh\ell}\right)^2}
\right].
\label{eq:maximum-odd-velocity}
\end{align}

\section{Asymptotic limits}
\label{sec:asymptotic-limits}

For narrow channels, $\ell\ll1$, the integral functions behave as
\begin{align}
\mathcal{A}_{\mathrm{e}}(\ell)
=
\frac{\ell^2}{3}-\frac{2\ell^4}{15}+O(\ell^6),
\,\,
\mathcal{A}_{\mathrm{o}}(\ell)
=
\frac{\ell^3}{24}-\frac{\ell^5}{240}+O(\ell^7).
\label{eq:narrow-channel-integral-asymptotics}
\end{align}
For wide channels, $\ell\gg1$,
\begin{align}
\mathcal{A}_{\mathrm{e}}(\ell)
=
1-\frac{1}{\ell}+O(e^{-2\ell}),
\,\,
\mathcal{A}_{\mathrm{o}}(\ell)
=
\frac{\ell}{2}-1+O(e^{-\ell}).
\label{eq:wide-channel-integral-asymptotics}
\end{align}
The physical mean velocity of the even mode is therefore
\begin{align}
\bar{v}_x
\simeq
\frac{8\alpha\mu_r e^2J_0^S}{3(\mu+\mu_r)\hbar\sigma_0}L^2
\quad
(L\ll\lambda),
\label{eq:narrow-channel-mean-velocity}
\end{align}
\begin{align}
\bar{v}_x
\simeq
v_0\left(1-\frac{\lambda}{L}\right)
\quad
(L\gg\lambda).
\label{eq:wide-channel-mean-velocity}
\end{align}

\section{Transient response after spin injection}
\label{sec:transient-response}

For a pure odd response, define
\begin{align}
u_{\mathrm{o}}(\zeta,s)
=
\frac{v_{x,\mathrm{o}}(y,t)}{V_*J_{\mathrm{o}}}.
\label{eq:dimensionless-odd-velocity}
\end{align}
It satisfies
\begin{align}
\frac{\partial u_{\mathrm{o}}}{\partial s}
=
\frac{\partial^2u_{\mathrm{o}}}{\partial\zeta^2}
-
\frac{\sinh\zeta}{\sinh\ell},
\,\,
u_{\mathrm{o}}(-\ell,s)=u_{\mathrm{o}}(\ell,s)=0,
\,\,
u_{\mathrm{o}}(\zeta,0)=0.
\label{eq:odd-dimensionless-pde}
\end{align}
The solution is
\begin{align}
u_{\mathrm{o}}(\zeta,s)
=
\mathcal{F}_{\mathrm{o}}(\zeta;\ell)
-
\sum_{n=1}^{\infty}
A_n^{\mathrm{o}}
\exp\left[-\left(\frac{n\pi}{\ell}\right)^2s\right]
\sin\left(\frac{n\pi\zeta}{\ell}\right),
\label{eq:odd-transient-solution}
\end{align}
where
\begin{align}
A_n^{\mathrm{o}}
=
\frac{2(-1)^n}{n\pi\left[1+(n\pi/\ell)^2\right]}.
\label{eq:odd-transient-coefficient}
\end{align}

For a pure even response, define
\begin{align}
u_{\mathrm{e}}(\zeta,s)
=
\frac{v_{x,\mathrm{e}}(y,t)}{V_*J_{\mathrm{e}}}.
\label{eq:dimensionless-even-velocity}
\end{align}
It satisfies
\begin{align}
\frac{\partial u_{\mathrm{e}}}{\partial s}
=
\frac{\partial^2u_{\mathrm{e}}}{\partial\zeta^2}
+
\frac{\cosh\zeta}{\cosh\ell},
\,\,
u_{\mathrm{e}}(-\ell,s)=u_{\mathrm{e}}(\ell,s)=0,
\,\,
u_{\mathrm{e}}(\zeta,0)=0.
\label{eq:even-dimensionless-pde}
\end{align}
The solution is
\begin{align}
u_{\mathrm{e}}(\zeta,s)
=
\mathcal{F}_{\mathrm{e}}(\zeta;\ell)
-
\sum_{n=0}^{\infty}
A_n^{\mathrm{e}}
\exp(-q_n^2s)\cos(q_n\zeta),
\label{eq:even-transient-solution}
\end{align}
with
\begin{align}
q_n
=
\frac{(2n+1)\pi}{2\ell},
\,\,
A_n^{\mathrm{e}}
=
\frac{4(-1)^n}{(2n+1)\pi(1+q_n^2)}.
\label{eq:even-transient-coefficient}
\end{align}
The general time-dependent response is obtained by linear superposition,
\begin{align}
v_x(y,t)
=
V_*
\left[J_{\mathrm{o}}u_{\mathrm{o}}(\zeta,s)+J_{\mathrm{e}}u_{\mathrm{e}}(\zeta,s)\right].
\label{eq:general-transient-response}
\end{align}
The slowest relaxation times are
\begin{align}
\tau_{\mathrm{o}}
=
\frac{L^2}{\pi^2\nu}
=
\frac{\ell^2}{\pi^2}\frac{\lambda^2}{\nu},
\,\,
\tau_{\mathrm{e}}
=
\frac{4L^2}{\pi^2\nu}
=
\frac{4\ell^2}{\pi^2}\frac{\lambda^2}{\nu}.
\label{eq:odd-even-relaxation-times}
\end{align}
For $L/\lambda=10$, Eq.~\ref{eq:odd-even-relaxation-times} yields
\begin{align}
\tau_{\mathrm{o}}
\simeq
10.1\frac{\lambda^2}{\nu},
\,\,
\tau_{\mathrm{e}}
\simeq
40.5\frac{\lambda^2}{\nu}.
\label{eq:relaxation-times-lambda-ten}
\end{align}
Thus profiles at $s=O(1)$ are still in the early hydrodynamic build-up regime, especially for the even pumping mode.

The time-dependent mean velocity of the even mode is
\begin{align}
\frac{\bar{v}_{x,\mathrm{e}}(t)}{V_*J_{\mathrm{e}}}
=
\mathcal{A}_{\mathrm{e}}(\ell)
-
\sum_{n=0}^{\infty}
\frac{8\exp(-q_n^2s)}{[(2n+1)\pi]^2(1+q_n^2)}.
\label{eq:time-dependent-even-mean-velocity}
\end{align}
The odd mode has zero cross-section average at all times,
\begin{align}
\int_{-L}^{L}v_{x,\mathrm{o}}(y,t)dy
=
0.
\label{eq:odd-zero-mean-all-times}
\end{align}

\section{Finite-frequency spin-mechanical admittance}
\label{sec:finite-frequency-admittance}

Let the boundary spin-current amplitude oscillate as
\begin{align}
J_{\mathrm{o,e}}(t)
=
J_{\mathrm{o,e}}(\omega)e^{-i\omega t}.
\label{eq:ac-boundary-current}
\end{align}
If the spin accumulation is quasistatic on the time scale of interest, the spatial profile in Eq.~\ref{eq:general-spin-potential} is unchanged and each hydrodynamic mode acquires a viscous response factor. For the odd mode,
\begin{align}
\frac{v_{x,\mathrm{o}}(y,\omega)}{V_*J_{\mathrm{o}}(\omega)}
=
\sum_{n=1}^{\infty}
\frac{A_n^{\mathrm{o}}\sin(n\pi\zeta/\ell)}{1-i\omega\tau_n^{\mathrm{o}}},
\,\,
\tau_n^{\mathrm{o}}
=
\frac{L^2}{\nu n^2\pi^2}.
\label{eq:odd-frequency-response}
\end{align}
For the even mode,
\begin{align}
\frac{v_{x,\mathrm{e}}(y,\omega)}{V_*J_{\mathrm{e}}(\omega)}
=
\sum_{n=0}^{\infty}
\frac{A_n^{\mathrm{e}}\cos(q_n\zeta)}{1-i\omega\tau_n^{\mathrm{e}}},
\,\,
\tau_n^{\mathrm{e}}
=
\frac{4L^2}{\nu(2n+1)^2\pi^2}.
\label{eq:even-frequency-response}
\end{align}
At $\omega=0$, Eqs.~\ref{eq:odd-frequency-response} and \ref{eq:even-frequency-response} reduce to the steady profiles in Eq.~\ref{eq:normalized-steady-profiles}. The even-mode spin-mechanical admittance for the mean flow is
\begin{align}
\mathcal{Y}_{\mathrm{e}}(\omega,L)
=
\frac{\bar{v}_{x,\mathrm{e}}(\omega)}{J_{\mathrm{e}}(\omega)}
=
V_*
\sum_{n=0}^{\infty}
\frac{8}{[(2n+1)\pi]^2(1+q_n^2)}
\frac{1}{1-i\omega\tau_n^{\mathrm{e}}}.
\label{eq:even-mean-flow-admittance}
\end{align}
The corresponding odd-mode orbital-angular-momentum admittance is
\begin{align}
\mathcal{Y}_{\mathcal{J}}(\omega,L)
=
\frac{\mathcal{J}_{z}^{\mathrm{orb}}(\omega)}{J_{\mathrm{o}}(\omega)}
=
4\rho V_*L^2
\sum_{n=1}^{\infty}
\frac{1}{(n\pi)^2\left[1+(n\pi/\ell)^2\right]}
\frac{1}{1-i\omega\tau_n^{\mathrm{o}}}.
\label{eq:orbital-angular-momentum-admittance}
\end{align}
At $\omega=0$, Eq.~\ref{eq:orbital-angular-momentum-admittance} reduces to Eq.~\ref{eq:orbital-angular-momentum-scaling}.

The characteristic viscous roll-off frequencies are set by
\begin{align}
\omega_{\nu,\mathrm{o}}
=
\frac{1}{\tau_{\mathrm{o}}}
=
\frac{\pi^2\nu}{L^2},
\,\,
\omega_{\nu,\mathrm{e}}
=
\frac{1}{\tau_{\mathrm{e}}}
=
\frac{\pi^2\nu}{4L^2}.
\label{eq:viscous-rolloff-frequencies}
\end{align}
Thus the finite-frequency mechanical response is controlled by the dimensionless parameter $\omega L^2/\nu$.

If the spin diffusion itself is dynamic, Eq.~\ref{eq:spin-diffusion-equation} is replaced by
\begin{align}
D_s\frac{d^2\mu_z^S}{dy^2}
-
\left(\frac{1}{\tau_s}-i\omega\right)\mu_z^S
=
0.
\label{eq:dynamic-spin-diffusion}
\end{align}
The spin-diffusion length becomes complex,
\begin{align}
\lambda_\omega
=
\sqrt{\frac{D_s}{\tau_s^{-1}-i\omega}}
=
\frac{\lambda}{\sqrt{1-i\omega\tau_s}},
\,\,
\lambda^2
=
D_s\tau_s.
\label{eq:complex-spin-diffusion-length}
\end{align}
At low frequencies, $\omega\tau_s\ll1$, the dc scaling functions derived above remain valid. At higher frequencies the replacement $\lambda\to\lambda_\omega$ determines the spin-diffusive part of the response.

Keeping the finite microrotation response of the micropolar rotational sector introduces a second possible roll-off. A minimal linear form is
\begin{align}
\left(1+\tau_R\partial_t-\xi^2\nabla^2\right)\Omega_i
=
\frac{1}{2}(\nabla\times\mathbf{v})_i+
\alpha\mu_i^S,
\label{eq:finite-microrotation-response}
\end{align}
where
\begin{align}
\tau_R
=
\frac{I}{4\mu_r},
\,\,
\xi^2
=
\frac{K_\Omega}{4\mu_r}.
\label{eq:microrotation-time-and-length}
\end{align}
Here $K_\Omega$ is the leading coefficient of the couple-stress sector. In Fourier variables,
\begin{align}
\Omega_i(\mathbf{q},\omega)
=
\frac{\frac{1}{2}i(\mathbf{q}\times\mathbf{v})_i+
\alpha\mu_i^S(\mathbf{q},\omega)}{1-i\omega\tau_R+\xi^2q^2}.
\label{eq:microrotation-fourier-response}
\end{align}
Equation~\ref{eq:microrotation-fourier-response} is the rotational-sector filter used for the ac extension. When the relevant time scales are resolved, the frequency dependence separates viscous momentum diffusion, spin diffusion, and microrotation relaxation.

\section{Pressure-constrained channel}
\label{sec:pressure-constrained-channel}

If the channel is closed in the $x$ direction, the even spin-driven response is converted into a pressure gradient rather than a net volume flow. Let
\begin{align}
G
=
-\frac{\partial p}{\partial x}.
\label{eq:pressure-gradient-definition}
\end{align}
The pressure-driven correction satisfying the no-slip condition is
\begin{align}
v_p(y)
=
\frac{G}{2\eta}\left(L^2-y^2\right).
\label{eq:pressure-driven-correction}
\end{align}
The zero-flow condition imposes
\begin{align}
0
=
\bar{v}_{x,\mathrm{spin}}+
\frac{GL^2}{3\eta},
\,\,
G
=
-\frac{3\eta}{L^2}\bar{v}_{x,\mathrm{spin}}.
\label{eq:zero-flow-pressure-gradient}
\end{align}
For a pure even EdH response,
\begin{align}
G
=
-\frac{3\eta V_*J_{\mathrm{e}}}{L^2}
\left(1-\frac{\tanh\ell}{\ell}\right).
\label{eq:edh-pressure-gradient-scaling}
\end{align}
Thus a closed channel measures the same scaling function as the open-channel mean velocity.

\section{Interface corrections and additional scaling variables}
\label{sec:interface-corrections}

The one-parameter scaling in Eq.~\ref{eq:general-steady-velocity} assumes ideal spin-current boundary conditions and no-slip hydrodynamic boundaries. A finite spin conductance $g_s$ leads to a mixed spin boundary condition,
\begin{align}
j_y^S(\pm L)
=
g_{s,\pm}\left[\mu_{\mathrm{Pt},\pm}^S-\mu_z^S(\pm L)\right].
\label{eq:mixed-spin-boundary-condition}
\end{align}
A finite hydrodynamic slip length $b$ modifies the no-slip condition. With such interface corrections, the normalized velocity profile has the schematic form
\begin{align}
\frac{v_x(y)}{V_*J_0^S}
=
\mathcal{G}\left(
\frac{y}{\lambda},
\frac{L}{\lambda},
\frac{b}{\lambda},
\frac{g_{s,+}\lambda}{C_s},
\frac{g_{s,-}\lambda}{C_s}
\right).
\label{eq:extended-scaling-form}
\end{align}
The ideal fixed-current result is recovered for $b/\lambda\to0$ when $J_\pm$ are the controlled boundary variables. If the Pt contacts are described instead by prescribed spin accumulations, finite $g_s$ produces spin backflow; large $g_{s,\pm}\lambda/C_s$ restores the ideal boundary-current limit.

The micropolar theory permits a rotational boundary condition through the couple stress,
\begin{align}
n_jC_{ji}
=
g_\Omega\left(\Omega_i-\Omega_i^{\mathrm{wall}}\right)+\tau_i^{\mathrm{surf}}.
\label{eq:rotational-boundary-condition}
\end{align}
These rotational boundary terms introduce additional dimensionless ratios, such as $K_\Omega/(g_\Omega\lambda)$ and $\xi/\lambda$. They are omitted in the ideal scaling law; different rotational anchoring conditions change them.

\section{Channel scales and viscous roll-off}
\label{sec:realistic-channel-scales}

No pressure-driven injection of liquid metal through a submicron channel is required. In the operating geometry, the liquid metal is pre-confined in a Pt/liquid-metal/Pt microchannel or in a sealed thin slab, and the spin-driven experiment then measures small-amplitude hydrodynamic motion of the sealed conductor.

The capillary pressure needed to push a liquid metal meniscus into a channel of effective radius $r$ is estimated by
\begin{align}
\Delta p_{\mathrm{cap}}
\sim
\frac{2\gamma}{r}.
\label{eq:capillary-pressure-estimate}
\end{align}
With a representative liquid-metal surface tension $\gamma\sim0.5\,\mathrm{N\,m^{-1}}$, this yields
\begin{align}
\Delta p_{\mathrm{cap}}
\sim
1\,\mathrm{MPa}
\left(\frac{\gamma}{0.5\,\mathrm{N\,m^{-1}}}\right)
\left(\frac{1\,\mu\mathrm{m}}{r}\right).
\label{eq:capillary-pressure-numerical}
\end{align}
Thus the experimentally relevant geometry is a pre-filled or pre-patterned Pt/liquid-metal/Pt channel with width $2L$ in the tens to hundreds of micrometers, or a sealed thin-film geometry when smaller $L$ is required.

The ac response is controlled by viscous momentum diffusion. For the even pumping mode,
\begin{align}
f_{\nu,\mathrm{e}}
=
\frac{1}{2\pi\tau_{\mathrm{e}}}
=
\frac{\pi\nu}{8L^2},
\label{eq:even-frequency-cutoff}
\end{align}
and the odd counterflow cutoff is four times larger. For a representative kinematic viscosity $\nu=3\times10^{-7}\,\mathrm{m^2s^{-1}}$,
\begin{align}
f_{\nu,\mathrm{e}}
\simeq
12\,\mathrm{Hz}
\left(\frac{\nu}{3\times10^{-7}\,\mathrm{m^2s^{-1}}}\right)
\left(\frac{100\,\mu\mathrm{m}}{L}\right)^2.
\label{eq:frequency-cutoff-estimate}
\end{align}
Thus $L=50$--$100\,\mu\mathrm{m}$ devices place the viscous roll-off in the Hz to 100 Hz range, while $L\sim10\,\mu\mathrm{m}$ sealed slabs can reach the kHz range. The measured roll-off and phase lag correspond to the spin-mechanical admittance in Eq.~\ref{eq:even-mean-flow-admittance}.

\section{Functions for plotting}
\label{sec:functions-for-plotting}

The universal steady profiles are
\begin{align}
\mathcal{F}_{\mathrm{o}}(\zeta;\ell)
=
\frac{\sinh\zeta}{\sinh\ell}
-
\frac{\zeta}{\ell},
\,\,
\mathcal{F}_{\mathrm{e}}(\zeta;\ell)
=
1-
\frac{\cosh\zeta}{\cosh\ell}.
\label{eq:plot-profile-functions}
\end{align}
The integral-observable functions are
\begin{align}
\mathcal{A}_{\mathrm{e}}(\ell)
=
1-
\frac{\tanh\ell}{\ell},
\,\,
\mathcal{A}_{\mathrm{o}}(\ell)
=
\frac{\ell}{2}
-
\tanh\frac{\ell}{2},
\label{eq:plot-integral-functions}
\end{align}
\begin{align}
\mathcal{J}_{\mathrm{o}}(\ell)
=
\frac{\ell^2}{3}+1-\ell\coth\ell.
\label{eq:plot-angular-momentum-function}
\end{align}
The maximum-velocity functions are
\begin{align}
\mathcal{M}_{\mathrm{e}}(\ell)
=
1-
\operatorname{sech}\ell,
\label{eq:plot-even-maximum-function}
\end{align}
\begin{align}
\mathcal{M}_{\mathrm{o}}(\ell)
=
\frac{1}{\ell}
\left[
\operatorname{arcosh}\left(\frac{\sinh\ell}{\ell}\right)
-
\sqrt{1-\left(\frac{\ell}{\sinh\ell}\right)^2}
\right].
\label{eq:plot-odd-maximum-function}
\end{align}
For transient profiles, use Eqs.~\ref{eq:odd-transient-solution} and \ref{eq:even-transient-solution}. For the even-mode ac mean-flow admittance, use Eq.~\ref{eq:even-mean-flow-admittance}.

\end{document}